\documentclass{eas}
\usepackage{graphicx}
\usepackage{natbib}
\usepackage{amssymb}
\def\kms{\,km\,s$^{-1}$~}     
\def\teff{$T_{\rm eff}$~} 
\def\vsini{$v\sin i$~} 
\begin{document}

\title{Lithium in field Am and normal A--F-type stars}

\runningtitle{P. North \etal: Li in Am and normal A--F stars}
\author{P. North}
\address{Laboratoire d'Astrophysique de l'Ecole Polytechnique F\'ed\'erale
de Lausanne (EPFL), Observatoire,
CH--1290 Sauverny, Switzerland}
%
\author{F. Betrix}
\address{Formerly at the (now obsolete) Institut d'Astronomie de
l'Universit\'e de Lausanne,
CH--1290 Chavennes-des-Bois, Switzerland}
\author{C. Besson}
\sameaddress{2}
%
%
\begin{abstract}
Preliminary abundances of lithium and a few other elements have been obtained
for 31 field Am stars with good Hipparcos parallaxes, as well as for
36 normal A and F stars. Radial and projected rotational velocities 
were determined as well. We examine the Li abundance as a function of
the stellar parameters: for normal stars, it is clearly bimodal for
\teff $< 7500$~K, while Am-Fm stars are all somewhat Li-deficient
in this range. The most Li-deficient stars -- either Am or normal -- tend
to be at least slightly evolved, but the reverse is not true.
\end{abstract}
\maketitle
\section{Introduction}
The abundance of Li has attracted much attention, especially since the Li gap
has been discovered in the Hyades for stars with \teff $\sim 6600$~K. In the
context of radiative diffusion, it is interesting to examine the atmospheric
abundance of lithium in stars where such a mechanism is known to be at work from
the abundances of other elements, such as calcium, i.e. in the Am stars.
Such studies have been carried out especially by Burkhart \& Coupry \citep{bc91}
and Burkhart et al. \citep{bcfg05}.
Their conclusion was that in general, the Li abundance
of Am stars is close to the cosmic value ($\log N(Li)\sim 3.0$ in the scale
where $\log N(H)= 12.0$), although a small proportion of them are deficient.
The latter seem in general to be either evolved stars or, as recently suggested
by Burkhart et al. \citep{bcfg05}, to lie on the red side of the Am domain,
among the $\rho$ Puppis--like stars.

In this poster, we present Li abundances obtained for 31 Am stars and 36 normal
A and F stars in the field, all having Hipparcos parallaxes. This sample had
been defined before the Hipparcos era, on purely photometric criteria, but with
the purpose of testing how far the Li abundance depends on the evolutionary
state, i.e. on the surface gravity $\log g$. The Hipparcos data which became
available later showed that the photometric luminosity calibrations of Am stars
were not very satisfactory (North et al. 1997), but allowed to determine
$\log g$ in a more fundamental way. Furthermore, the sample has the advantage
of presenting no bias against large rotational velocities.

\section{Observations and analysis}
All stars were observed at OHP with the Aur\'elie spectrograph attached to the
1.5m telescope, in April 1993 and in October 1993 and 1994. The grating No 7 was
used, giving a resolving power $R=40000$ in the spectral range $6640-6760$~\AA .
The typical exposure times were between 40 and 60 minutes, the resulting
signal-to-noise ratio being between 250 and 400. The spectra were
reduced during the observing runs with the IHAP package, and were later
normalized to the continuum in an interactive way.

The analysis was made by comparison of the observed spectra with synthetic ones
convoluted with an assumed gaussian instrumental profile and with an appropriate
rotational profile. The Synspec code (Hubeny et al. 1995) and Kurucz model
atmospheres were used to produce the synthetic spectra. The line parameters were
taken from Kurucz's $gfiron$ list, except of course the parameters for the Li
doublet. The effective temperatures were computed from Geneva photometry, while
the surface gravities were computed from the Hipparcos parallaxes, by combining
them with theoretical evolutionary tracks from Schaller et al. \citep{ssmm92},
as explained by North \citep{n98}, assuming standard evolution.
The microturbulent velocity was either computed from the formula proposed by
Edvardsson et al. \citep[eqn 9]{e93}, for \teff $< 7000$~K, or estimated from the
Fig.~1 of Coupry \& Burkhart \citep{cb92}, for \teff $\geq 7000$~K.
The abundance of
Fe, Ca and a few other elements (in cases of sharp lined stars) were first
estimated by visual fits. Then, the Li abundance and the projected rotational
velocities were obtained by minimizing the $\chi^2$ between observed and
synthetic spectra having various values of these parameters.
\begin{figure}
  \includegraphics[width=10cm]{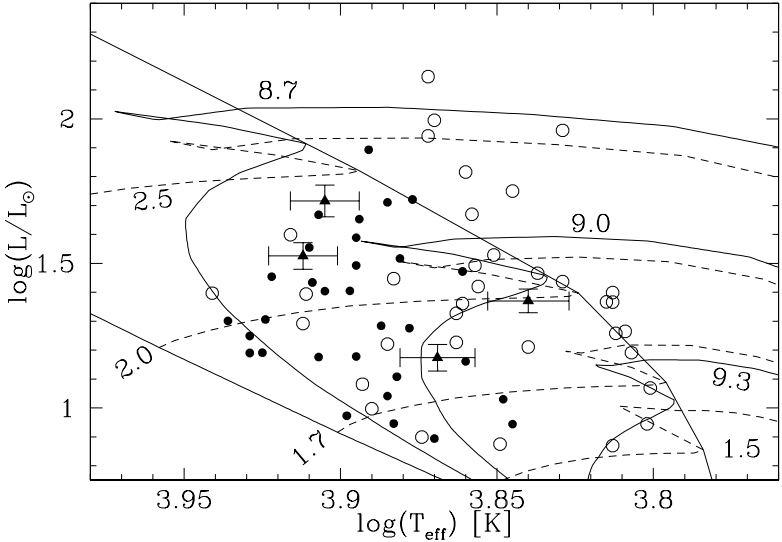}
  \caption{HR diagram of the Am (black dots) and normal (white dots) stars of
our sample. Black triangles (with error bars typical of the whole
sample) are stars from Burkhart et al. (2005, Table~3) not in
common with our sample. The error bars were drawn assuming a $\pm 200$~K error
on \teff and include, on the vertical axis, the parallax error of Hipparcos.
}
\label{hr}
\end{figure}
\section{Results}
Fig.~\ref{hr} shows the distribution of Am stars (full dots) and of normal A-F stars
(open dots) in the HR diagram. Evolutionary tracks and isochrones from
Schaller et al. \citep{ssmm92} are shown for 4 masses ($1.5$ to $2.5~M_\odot$)
and for 3
ages ($\log t = 8.7$ to $9.3$) respectively. The stars are well distributed
on the whole main sequence band. The lack of Am stars below \teff $\sim 7000$~K
is the well-known limit due to the onset of convection.

\begin{figure}
  \includegraphics[width=10.3cm]{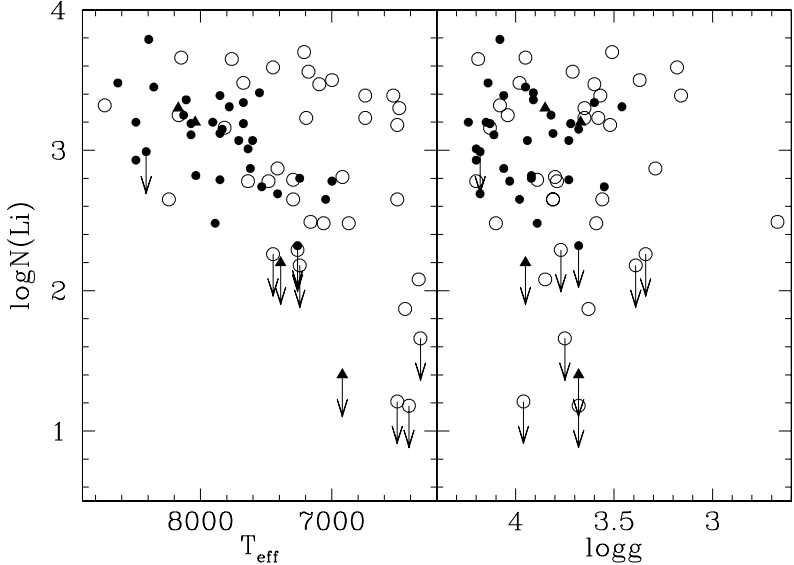}
  \caption{{\bf Left:} Li abundance (on the scale $\log N(H)=12$) of Am
stars (black
dots) and normal A--F-type stars (white dots) versus effective temperature.
Upper limits to the Li abundance are indicated by vertical arrows.
Black triangles are from Burkhart et al. \citep{bcfg05}.
{\bf Right:} Li abundance versus surface gravity derived from Hipparcos
parallaxes. The leftmost arrow refers to the Am star HD 18769, which has
\teff $=8420$~K and \vsini $= 46$~\kms, so that only an upper limit
to its Li abundance can be obtained. The typical error on $\log g$ is $0.1$~dex,
while that on $\log N(Li)$ vary from better than $0.1$~dex to more than
$0.3$~dex, depending on \teff and \vsini.}
\label{LiTelg}
\end{figure}

Fig.~\ref{LiTelg} (left) shows the lithium abundance as a function of \teff
for Am stars (full dots) and for normal A--F stars (open dots).
The most striking feature of this diagram is
the bimodal distribution of the Li abundance for \teff $\lesssim 7500$~K, which
is reminiscent of a similar distribution of F-type dwarfs in the range $5900
<$ \teff $< 6600$~K reported by Lambert et al. \citep[Fig.~4]{lhe91}.
Thus, our data
complement that of Lambert et al. as well as the larger sample of Chen et al.
\citep{cnbz01} by extending the results to higher \teff. We have
verified that duplicity cannot account for the low apparent Li abundances (even
though this might hold for some isolated cases). Restricting the diagram to
those stars with
\vsini $< 80$~\kms, the upper branch almost disappears (there are only two
normal stars left around \teff $\sim 6500$~K), while the lower one remains
intact. This is related to the fact that the upper branch is populated only with
normal stars, which rotate more rapidly than the Am stars, while the lower
branch is a mix of normal and Am stars. Thus, below $7500$~K, all Am stars of
our sample are Li deficient. The black triangles refer to the 4 stars of
Burkhart et al. \citep[their Table~3]{bcfg05} which are not common to our
sample. Their positions are in perfect agreement with the general picture.

Fig.~\ref{LiTelg} (right) displays the Li abundance as a function of surface gravity. There is
no strong trend, but one can notice that those stars (either Am or normal) which
are strongly deficient in Li are {\bf all} at least slightly evolved. There is one
unevolved star (HD 18769) for which only an upper limit to its Li abundance
could be obtained, but this is due to its high \teff ($8420$~K) and moderately
broad lines, and the upper limit is close to the ``cosmic'' Li abundance, so
this is not a significant exception. Thus, we confirm the suggestion made by
Burkhart \& Coupry that Li-deficient Am stars are evolved objects, although it
seems that all evolved Am stars are not necessarily deficient.
%


\end{document}